\documentclass[a4paper,11pt]{article}
\usepackage[utf8]{inputenc}
\usepackage{amsfonts}
\usepackage{amssymb}
\usepackage{graphicx}
\usepackage{fancyhdr}
\usepackage{fancyvrb}
\usepackage{fancybox}
\usepackage{float}
\usepackage{geometry}
\usepackage{minitoc}
 \usepackage{indentfirst}
 \usepackage{multicol}
 \usepackage[outerbars]{changebar}
 \usepackage{pifont,textcomp}
 \usepackage{amsfonts}
 \usepackage{amsmath}
 \usepackage{amscd}
 \usepackage{array}
\usepackage{multirow}
 \usepackage{mathrsfs}
 \usepackage{amssymb}
 \usepackage{amsthm}

\title{Minimal and maximal lengths from position-dependent noncommutativity}
\author{Latévi M. Lawson\\
\space\\
 Université de Lomé, Faculté des Sciences, Departement de Physique,\\
Laboratoire de Physique des Matériaux et des Composants\\
à Semi-Conducteurs, 01 BP 1515 Lomé, Togo\\\\
lawmenx@gmail.com}

\begin{document}
\maketitle

\begin{abstract}
Fring {\it et al}. in Ref.\cite{1} have introduced a new set 
of noncommutative space-time commutation relations in two space dimensions.  
It had been  shown that any fundamental objects introduced in this space-space 
noncommutativity   are string-like. Taking this result into account, we generalize the seminal work of Fring {\it et al}
to the case that there is also a maximal  length   from position-dependent  noncommutativity    and  a minimal momentum  
arising  from generalized versions of Heisenberg’s uncertainty relations. The existence of maximal length is related 
to the presence of  an extra, first order term in particle’s length  that provides the basic difference of our analysis with theirs. This maximal
length breaks up  the  well known singularity problem of  space time. We establish different representations  of this noncommutative space and finally
 we study some basic and interesting quantum mechanical systems in these new variables.

\end{abstract}


\section{Introduction}
One of the oldest open problems in modern physics is the unification of General Relativity (GR)
and Quantum Theory (QT). The problem of finding a quantum formulation of the Einstein equation in
GR still does not have a consistent and satisfactory solution.
The difficulty arises since GR deals with the events which define the world-lines of particles, 
while quantum mechanics do not allow the definition of trajectory. 
Nevertheless, one of the most active candidate theories to address this problem, string theory, 
predicted that this unification should occur at the Planck scale and should give birth to quantum gravity \cite{2,3}.
Thus, the minimal measurement of quantum gravity indicates  a  measurement  of Planck order $l_p = 10^{-35}m$. 
This value is extremely small; its experimental search lies beyond the energies currently accessible in the laboratory. 
In the theoretical framework, the observational search for such existence of a minimal length can be derived from the so called Generalized
(Gravitational) Uncertainty Principle (GUP)\cite{4}
\begin{eqnarray}
\Delta x\Delta p\geq \frac{\hbar}{2}\left[\mathbb{I}+\beta (\Delta p)^2\cdots\right],
\end{eqnarray}
by deforming the Heisenberg algebra as follows
\begin{eqnarray}
[\hat x,\hat p]=i\hbar(\mathbb{I}+\beta \hat p^2\cdots).
\end{eqnarray}
This latter implies a minimal position uncertainty $\Delta x_{min}$ \cite{4,5,6,7,8}. Moreover, the emergence of this minimal length in non-relativistics quantum mechanics introduces many consequences such as the deformation of the Heisenberg algebra, the loss of the localization of particles in the position representation, the deformation of the structures of the Hilbert space, the noncommutation in position space \cite{4} etc. In quantum geometry as in quantum gravity, this minimal length induces an addition to the
previous consequences observed in the Hilbert space, the violation of the Lorentz invariance \cite{9,10} 
and an intriguing mixing between the Ultraviolet and the Infrared \cite{11}. It  leads to a generalized  Hawking temperature 
\cite{12,13} and removal of the Chandrasekhar limit in cosmology \cite{14} etc.

Since the appearance in quantum mechanics, 
many alternative approaches to improve this minimal length had been introduced \cite{15,16,17,18} which propose higher modifications to GUP.
In this sense, a new set of noncommutative space-time commutation relations in two dimensional configuration space has been
recently introduced \cite{1}. The space-space commutation relations are deformations of the standard flat noncommutative space-time relations that have position dependent structure constants. These deformations lead to minimal lengths and it has been found that 
any object in this two dimensional space is string-like, in the sense that having a fundamental length beyond
which a resolution is impossible. Some extensions of this work have been done in \cite{1,19,20,21,22} and the model of 
gravitational quantum well have been solved in these new variables \cite{23}.

In this paper, we are  going to generalize the result  of
Fring {\it et al.}\cite{1} to the case that the existence of a  maximal length is considered too. In  this seminal work, one notes that a 
simultaneous measurement in  position space-time  leads to a minimal length in $X$-direction  as well as a minimal length
in $Y$-direction  when  informations are given-up in one of the  directions. Here we just consider the case  where for  a simultaneous measurement, 
the lost of  particle's localization in  $X$-direction leads to its  maximal localization in $Y$-direction.
Then, both  minimal momentum  and maximal length arise from  the generalized  versions of Heisenberg’s uncertainty relations 
for a simultaneous measurement  in $Y,P_y$-directions.
This proposal agrees with a similar  perturbative approaches predicted by  Doubly Special Relativity theories (DSR) \cite{24,25} 
and by the seminal result of  Nozari and Etemadi \cite{26}.  The existence of maximal length  related 
to the presence of  an extra, first order term in particle’s length, brings a lot of new features to the Hilbert space
representation of quantum mechanics at the Planck scale. Moreover, the presence of minimal uncertainties  in the representation of
this  algebra,  allows us to work  with the  position   $ Y$-space representation. In this manner, we  explore the quantum physical 
implications and Hilbert space representation in the presence of  minimal measurable uncertainties and a maximal measurement length. 
Eventually, in order to  avoid  the  ambiguity of the meaning of  wavefunction due to the existence of  minimal measurable uncertainties,
we propose  another  representation of operators  $\hat X, \hat Y, \hat P_x, \hat P_y$ in terms of  
standard Heisenberg operators   $\hat x_s, \hat y_s, \hat p_{x_s}, \hat p_{y_s}$
through approximations in first order of deformed parameters $\theta,\tau$. 
This realization  makes an  
effective noncommutative space and the  whole phase space structure of the Lie-algebraic type  is related to $\kappa$-like deformations
of space and deformed Heisenberg algebra \cite{27,28,29,30,31}.
 
 In the present paper we   study some interesting  quantum mechanics systems
  in two-dimensional position dependent  noncommutative spaces 
 and we determine how the Schr\"{o}dinger equation  in the  reduced 
 noncommutative algebra  can be solved exactly or  perturbatively.
  The paper is organized as follows. In section (\ref{sec2}), we review the  Heisenberg algebra and its deformation in two-dimensional
  quantum mechanics with theirs corresponding consequences as we have  recently introduced
\cite{23}. In section (\ref{sec3}), we introduce the new set  of position-dependent noncommutative space and 
we derive  minimal uncertainties and a maximal length   resulting from this space and the representations of wavefunction.
In section (\ref{sec4}), we study 
some simple models formulated in terms of our new set of variables such as the free particle, the particle in a box and the harmonic oscillator.
The conclusion is given  in section $(\ref{section5})$.

\section{ Heisenberg algebra and its deformation}\label{sec2}
 Let $\mathcal{H}=\mathcal{L}^2(\mathbb{R}^d)$ be 
the  Hilbert space  of square integrable functions $\psi(x)$ in $d$-dimensional Euclidian space.
The scalar product on $\mathcal{H}$ is defined
\begin{equation}
\langle \phi|\psi\rangle= \int_{\mathbb{R}^d} d^d x\phi^*(x)\psi(x).
\end{equation}
We denote the elements of this Hilbert space by
$\psi(x)\equiv |\psi\rangle$ and the elements   of its dual by $\langle \psi|$, 
which maps elements of $\mathcal{L}^2(\mathbb{R}^d)$ onto complex numbers by $\langle \psi|\phi\rangle=(\psi|\phi)$. 
The corresponding norm is given as usual by $||\psi||=\sqrt{\langle \psi|\psi\rangle}$  \cite{32}.
Let also consider  a physical observable represented by  a 
Hermitian operator   $\hat A$ defined on its domain $\mathcal{D} (\hat A)$ 
maximal dense on
$\mathcal{H}$  and  $\hat A^\dag$  its  adjoint  defined on  $ \mathcal{D} (\hat A^\dag )$  such as   
 \begin{equation}
  \langle\phi | \hat A\psi\rangle = \langle \hat A^\dag  \phi  |\psi\rangle,
 \end{equation}
  where  $|\phi\rangle\in  \mathcal{D} (\hat A^\dag )$  and  $|\psi\rangle\in  \mathcal{D} (\hat A)$.
  The fact  that $\hat A=\hat A^\dag$ 
 ensures   the expectation value   $  \langle\psi | \hat A|\psi\rangle$ is real, the
inner products of wavefunctions in $\mathcal{H}$  have a positive norm and that the time evolution operator is unitary. This situation does not 
prove  that $\hat A$ is  truly self-adjoint because in general the domains $ \mathcal{D} (\hat A)$ and $\mathcal{D} (\hat A^\dag)$ may be different.
Therefore, the self-adjointness of $\hat A$  results from the fact that   $\mathcal{D} (\hat A )= \mathcal{D} (\hat A^\dag )$ and  
 $\hat A=\hat A^\dag$. For  simultaneous measurement of two observables $\hat A $ and $\hat B$ in the  state $|\psi\rangle$, the uncertainty
satisfies the inequality 
\begin{eqnarray}\label{incertitude}
 \Delta A\Delta B\geq \frac{\hbar}{2}\left|\langle \psi |[\hat A,\hat B]|\psi\rangle\right|,
\end{eqnarray}
 where  $\Delta A$  and $ \Delta B$ are respectively, the dispersions defined as 
 $\Delta A^2:=\langle \psi| \hat A^2|\psi\rangle  -\langle \psi |\hat A |\psi \rangle^2$   and 
$\Delta B^2:=\langle \psi| \hat B^2|\psi\rangle  -\langle \psi |\hat B |\psi \rangle^2$. From the equation  (\ref{incertitude}), we deduce the following
relation, that is
\begin{eqnarray}
 \left|\left| \left( \hat A- \langle \hat A\rangle  +\frac{\langle [\hat A,\hat B]\rangle }{2\Delta B^2 }
 \left(\hat B-\langle \hat B \rangle \right)
|\psi\rangle
 \right) \right|\right|\geq 0.
\end{eqnarray}
The Fourier transform of the   wavefunction $\psi(x)$ is 
denoted  by  $\psi(p)$ with $p\in \mathbb{R}^d $   is given by
\begin{eqnarray}\label{Four}
 \psi(p)=\frac{1}{(2\pi\hbar)^{3/2} }\int_{\mathbb{R}^d} \psi(x) e^{-\frac{i}{\hbar}p.x} d^dx,
\end{eqnarray}
 and the inverse transform is given by
 \begin{eqnarray}
  \psi(x)=\frac{1}{(2\pi\hbar)^{3/2} }\int_{\mathbb{R}^d} \psi(p) e^{\frac{i}{\hbar}p.x} d^dp.
 \end{eqnarray}

Now, let start with the  following  definition:\\
{\bf Definition 2.1.} 
 {\it In  $d$-dimensional space, a unitary representation of  the Heisenberg algebra is}
\begin{eqnarray}
 [\hat x_i,\hat x_j]=0,\quad [\hat x_i,\hat p_j]= i\hbar \delta_{ij},\quad 
[\hat p_i,\hat  p_j]=0,\quad i,j=1,2\cdots d,
\end{eqnarray} 
{\it where  $\hat x_i$ and $\hat p_j$  are Hermitian operators acting on $\mathcal{H}$}.

In $2$-dimensions of this algebra, we have :\\
{\bf Proposition 2.1.} {\it Let $\mathcal{H}_s=\mathcal{L}^2(\mathbb{R}^2)$  be the Hilbert space that defined the algebra of linear  operators  in 2D  commutative  space }
\begin{eqnarray}\label{alg1}
 [\hat x_s,\hat y_s]&=&0,\,\,\,[\hat x_s,\hat p_{x_s}]=i\hbar,\,\,\,[\hat y_s,\hat  p_{y_s}]=i\hbar, \cr
[\hat p_{x_s},\hat  p_{y_s}]&=&0,\,\,\, [\hat x_s,\hat  p_{y_s}]=0,\,\,\,\,\,\,\, [\hat y_s, \hat p_{x_s}]=0.
\end{eqnarray}
{\it where the  operators $\hat x_s,\hat y_s,\hat p_{x_s},\hat  p_{y_s}$ are Hermitian operators acting on the space of 
square integrable function of $\mathcal{H}_s$}.\\
These commutation relations  lead to the standard uncertainty relations
\begin{eqnarray}
 \Delta  x_s\Delta p_{x_s}\geq\frac{\hbar}{2},\,\,\,
 \Delta  y_s\Delta  p_{y_s}\geq\frac{\hbar}{2}.
\end{eqnarray}
Consequently,  the Schrödinger representation  of the algebra in (\ref{alg1}) is 
\begin{eqnarray}
 \hat x_s \psi(x_s,y_s)&=&x_s.\psi(x_s,y_s),\,\,\,
 \hat y_s \psi(x_s,y_s)=y_s.\psi(x_s,y_s),\\
  \hat p_{x_s} \psi(x_s,y_s)&=&-i\hbar\frac{\partial}{\partial x_s}\psi(x_s,y_s),\,\,\,
  \hat p_{y_s} \psi(x_s,y_s)=-i\hbar\frac{\partial}{\partial y_s}\psi(x_s,y_s),
\end{eqnarray}
where $\psi(x_s,y_s)\in \mathcal{H}_s$.
The above $2D$ Heisenberg algebra will be now replaced by the non-commutative Heisenberg
algebra.

{\bf Proposition 2.2.} {\it Let $\mathcal{H}_0=\mathcal{L}^2(\mathbb{R}^2)$  be the Hilbert space that  describes the   
ordinary $2D$ noncommutative space. The Hermitian operators that act on this space satisfy the following relations
\begin{eqnarray}\label{alg2}
 [\hat x_0,\hat y_0]&=&i\theta,\,\,\,[\hat x_0,\hat p_{x_0}]=i\hbar,\,\,\,[\hat y_0,\hat p_{y_0}]=i\hbar,\cr
 {[\hat p_{x_0},\hat p_{y_0}]}&=&0,\,\,\quad [\hat x_0,\hat p_{y_0}]=0,\,\,\, [\hat y_0, \hat p_{x_0}]=0,
\end{eqnarray}
where $\theta\in \mathbb{R}_+^*$, is the noncommutative parameter which has the length square dimension. If $\theta$ is set to zero,
we obtain the standard Heisenberg commutations relations (\ref{alg1})}.

The noncommutation relations (\ref{alg2}) lead to an additional uncertainty due to the noncommutativity of the position operators
\begin{eqnarray}
 \Delta  x_0\Delta  y_0\geq\frac{|\theta|}{2},\,\, \Delta  x_0\Delta p_{x_0}\geq\frac{\hbar}{2},\,\,\,\,
 \Delta  y_0\Delta  p_{y_0}\geq\frac{\hbar}{2}.
\end{eqnarray}
Based on the fact that $\theta$ has dimension 
of (length)$^2$, then $\sqrt{\theta}$ defines a fundamental scale of length which 
characterizes the minimum uncertainty possible to achieve in measuring this quantity.

The action of these operators on the square integrable wavefunctions 
$\psi(x_0,y_0)\in \mathcal{H}_0$  can be realized  as follows
\begin{eqnarray}\label{rep2}
 \hat x_0 \psi(x_0,y_0)&=&x_0\star\psi(x_0,y_0),\,\,\, \hat y_0 \psi(x_0,y_0)=y_0\star\psi(x_0,y_0),\cr
  \hat p_{x_0} \psi(x_0,y_0)&=&-i\hbar\frac{\partial}{\partial x_0}\psi(x_0,y_0),\,\,\,
  \hat p_{y_0} \psi(x_0,y_0)=-i\hbar\frac{\partial}{\partial y_0}\psi(x_0,y_0),\,\,\,
\end{eqnarray}
where $\star$ denotes the so-called star product, defined by
\begin{equation}
 (f\star g)(x,y)=\exp\left(\frac{i}{2}\theta_{ij}\partial_{x_i}\partial_{y_j}\right)f(x)g(y),
\end{equation}
where $f$ and $g$ are two arbitrary infinitely differentiable functions on $\mathbb{R}^2$ 
is real and antisymmetric i.e $\theta_{ij}=\epsilon_{ij}\theta $
( $\epsilon_{ij}$ a completely antisymmetric tensor with $\epsilon_{1,2}=1$).

One possible way of implementing algebra Eqs.(\ref{alg2}) is to construct the 
noncommutative operators  $ \{\hat x_0,  \hat y_0, \hat p_{x_0}, \hat p_{y_0}\}$ from the commutative operators 
$ \{\hat x_s,  \hat y_s, \hat p_{x_s}, \hat p_{y_s}\}$ by means of a linear transformation namely Bopp-shift denoted by $\mathcal{B}_\theta$.
In the literature \cite{18,28}, there are many versions of the Bopp-shift such as  the asymmetric Bopp-shift
  \begin{eqnarray}\label{bob1}
\mathcal{B}_\theta^{a_1}:
\begin{cases}
 \hat x_0=\hat x_s-\frac{\theta}{2\hbar}\hat  p_{y_s},\\
 \hat y_0=\hat y_s,\\
 \hat p_{x_0}=\hat  p_{x_s},\\
 \hat p_{y_0}=\hat  p_{y_s},\\
 \end{cases}
 \quad \mbox{or} \quad
\mathcal{B}_\theta^{a_2}: \begin{cases}
 \hat x_0=\hat x_s,\\
 \hat y_0=\hat y_s+\frac{\theta}{2\hbar}\hat  p_{x_s}\\
 \hat p_{x_0}=\hat  p_{y_s},\\
 \hat p_{y_0}=\hat  p_{y_s},\\
 \end{cases}
\end{eqnarray}
and the symmetric Bopp-shift

 \begin{eqnarray}\label{bob}
\mathcal{B}_\theta^s:
\begin{cases}
 \hat x_0=\hat x_s-\frac{\theta}{2\hbar}\hat  p_{y_s},\\
 \hat y_0=\hat y_s+\frac{\theta}{2\hbar}\hat  p_{x_s}\\
 \hat p_{x_0}=\hat  p_{y_s},\\
 \hat p_{y_0}=\hat  p_{y_s}.\\
 \end{cases}
\end{eqnarray}
There are some advantages in using the asymmetric Bopp shift such as the decoupling of the
operators in some of the problems and some simplifications of expressions. In fact $\mathcal{B}_\theta^{a_1}$ and 
$ \mathcal{B}_\theta^{a_2}$ do not always lead to the same results for the same problems. For that reason the symmetrical Bopp shift 
$ \mathcal{B}_\theta^s$ is
often used \cite{33}. In the present work,
some of these transformations  will be used in the forthcoming developpement according to our purposes. Let remarks that 
with the transformations (\ref{bob1}) and (\ref{bob}), it is easy to verify that the operators $\hat x_0, \hat y_0 , \hat p_{x_0}, \hat p_{y_0}$
are Hermitian as we  mentioned in proposition 2.2.
Taking  the transformation  $\mathcal{B}_\theta^s$ for example , one changes in the Schr\"{o}dinger's representations (\ref{rep2}),
the star product by the usual  product of field  such as
\begin{eqnarray}
 \hat x_0 \psi(x_s,y_s)&=&x_s.\psi(x_s,y_s)+\frac{i\theta}{2}\frac{\partial}{\partial  y_s}\psi(x_s,y_s)
; \, \hat p_{x_0} \psi(x_s,y_s)=-i\hbar\frac{\partial}{\partial x_s}\psi(x_s,y_s),\label{15}\\ 
 \hat y_0 \psi(x_s,y_s)&=&y_s.\psi(x_s,y_s)-\frac{i\theta}{2}\frac{\partial}{\partial y_s}\psi(x_s,y_s);\,
  \hat p_{y_0} \psi(x_s,y_s)=-i\hbar\frac{\partial}{\partial y_s}\psi(x_s,y_s)\label{16}.
\end{eqnarray}
 The equations (\ref{15}) and (\ref{16}) are a realization for the deformed Heisenberg algebra in the case of Moyal noncommutativity.

\section{Measurement lengths from position dependent noncommutative space } \label{sec3}
\subsection{Position dependent noncommutative algebra  and uncertainty measurements }
This section addresses the construction of a  new set  of   noncommutative space by introducing  new  operators 
$\hat X,\hat Y, \hat P_x,\hat P_y$  and to convert the constant $\theta$ of 
the algebra (\ref{alg2}) into a fonction 
$\theta (X,Y)=\theta (1-\tau\hat Y +\tau^2 \hat Y^2)$. We start with the following proposition.\\
 
 {\bf Proposition 3.1.}
 {\it  Given new set of  Hermitian  operators  $\hat X,\hat Y, \hat P_x,\hat P_y$  defined on
 $\mathcal{H}_k=\mathcal{L}^2(\mathbb{R}^2)$ satisfy the following commutations relations and all possible permutations of the Jacobi identities }
\begin{eqnarray} \label{alg3}
[\hat X,\hat Y]&=&i\theta (1-\tau\hat Y +\tau^2 \hat Y^2),\quad [\hat X,\hat P_x ]=i\hbar (1-\tau \hat Y +\tau^2 \hat Y^2),\cr
 {[\hat Y,\hat P_y ]}&=&i\hbar (1-\tau \hat Y +\tau^2 \hat Y^2),\quad\quad
 { [\hat P_x,\hat P_y]}=0, \cr 
 {[\hat Y ,\hat P_x]}&=&0,\quad [\hat X,\hat P_y]=i\hbar\tau(2\tau \hat Y\hat X-\hat X)+
  i\theta\tau(2\tau \hat Y\hat P_y-\hat P_y),
\end{eqnarray}
{\it where $\tau\in\mathbb{R}_+^*$ is the  deformed parameter.
 By taking $\tau\rightarrow 0$,  we obviously recover the algebra (\ref{alg2}).}\\\\
 {\it Proof:} One  can recover  the algebra  (\ref{alg3}) by setting these operators  
 in terms of the  Hermitian operators $\hat x_0,\hat y_0, \hat p_{x_0},\hat p_{y_0}$  by using the following representation
  \begin{eqnarray}\label{reR}
   \mathcal{R}_\tau:
   \begin{cases}
    \hat X=\hat x_0-\tau \hat y_0\hat x_0+\tau^2\hat y_0^2\hat x_0,\\
    \hat Y=\hat y_0,\\
     \hat P_x=\hat p_{x_0},\\
     \hat P_y=\hat p_{y_0}-\tau \hat y_0\hat p_{y_0}+\tau^2\hat y_0^2\hat p_{y_0}.
     \end{cases}
  \end{eqnarray}
See Appendix  for the prove of  all possible permutations of the Jacobi identities.

  The     parameter $\tau$ can be compared to the deformed parameter $\beta =\frac{l_p^2}{\hbar^2}$ of \cite{4,26} such as $\Delta x=\hbar \sqrt{\beta}$, the minimal length of quantum gravity below which spacetime
distances cannot be resolved as predicted by string theory \cite{2}. Such a feature is  expected to be a candidate
 theory of quantum gravity, since gravity itself is characterized by the Planck length $l_p$. In the present case this parameter manifests
  as deformation of  the noncommutative space (\ref{alg2}) by  quantum gravity. 
The proposal (\ref{alg3}) is consistent with the similar prediction of  DSR \cite{24,25} 
and by the seminal result of  Nozari and Etemadi \cite{26}. From the representation (\ref{reR}), on can interpret 
 $\hat x_0, \hat y_0 , \hat p_{x_0}, \hat p_{y_0}$ as the set of operators at low energies which has the standard 
representation in position space and $\hat X,\hat Y, \hat P_x,\hat P_y $  as the set of operators at high energies, where they
have the generalized representation in position space.

 In comparison with the  Fring \textit{et al} noncommutative space  \cite{1}, here there is
an extra, first order term in particle’s length which will be the origin of  the existence of a maximal length. 
The presence of this 
term is the source of differences between  our set of algebra representation (\ref{alg3})
and Fring \textit{et al}'s algebra \cite{1}.                
   From  these commutation relations  (\ref{alg3}), 
an interesting features  can be observed through the  following  uncertainty relations:
\begin{eqnarray}
 \Delta  X\Delta Y&\geq&\frac{|\theta|}{2}\left(1-\tau\langle \hat Y\rangle+\tau^2\langle \hat Y^2\rangle\right),\label{in1}\\
 \Delta  Y\Delta P_y&\geq& \frac{\hbar}{2}\left(1-\tau\langle \hat Y\rangle+\tau^2\langle \hat Y^2\rangle\right),\label{in2}\\
 \Delta  X\Delta P_x&\geq& \frac{\hbar}{2}\left(1-\tau\langle \hat Y\rangle+\tau^2\langle \hat Y^2\rangle\right)\label{in3}.
 \end{eqnarray}

$i)$  In the situation of uncertainty relation (\ref{in1}), using 
$ \langle \hat Y^2\rangle=\Delta  Y^2+\langle \hat Y\rangle^2$,  this relation can be
rewritten  as a  second order equation for $\Delta Y$. The solution for $\Delta Y$ are as follows 
\begin{equation}
 \Delta Y=\frac{\Delta X}{\theta \tau^2}\pm \sqrt{\left(\frac{\Delta X}{\theta \tau^2}\right)^2
 -\frac{\langle \hat Y\rangle}{\tau}\left(\tau\langle \hat Y\rangle-1\right)
 -\frac{1}{\tau^2}}.
\end{equation}
The reality of solutions gives the following minimum
value for $\Delta X$
\begin{eqnarray}
\Delta X_{min}=\theta \tau\sqrt{1-\tau \langle \hat Y\rangle+\tau^2\langle \hat Y\rangle^2}.
\end{eqnarray}
Therefore, these equations lead to the absolute minimal uncertainty $\Delta X_{min}^{abs}$ in $X$ direction  and the absolute maximal uncertainty  $\Delta Y_{max}^{abs}$ in $Y$ direction  
for $\langle  \hat Y\rangle=0$, such as:
\begin{eqnarray} 
 \Delta X_{min}^{abs}&=&\theta\tau,\\
 \Delta Y_{max}^{abs}&=& l_{max}=\frac{1}{\tau}.\label{max}
\end{eqnarray}
In comparison with Fring \textit{et al}.   formalism \cite{1},
where a simultaneous measurement in  $X$ and $Y$ spaces leads to a minimal length for $\hat X$ or for 
$\hat Y$ when  informations are given-up in one direction, here a simultaneous measurement leads to a minimal measurement in $\hat X$ which introduces a lost of localization in $X$-direction and a maximal measurement in $\hat Y$ which conversely allows maximal localization in $Y$-direction.\\\\ 
ii) Repeating the same calculation  and argumentation in the situation of  uncertainty relation (\ref{in2}) for simultaneous $\hat Y,\hat P_y$-measurement, we find
the absolute  maximal uncertainty  $\Delta Y_{max}^{abs}$ (\ref{max}) and an absolute minimal uncertainty momentum $\Delta P_{y_{min}}^{abs}$  for  $\langle  \hat Y\rangle=0$, such 
\begin{equation}\label{minmom}
 \Delta P_{y_{min}}^{abs}=\hbar \tau.
\end{equation}
iii) Finally, for the uncertainty relation (\ref{in3}), a simultaneous $ \hat X, \hat P_{x}$-measurement
does not present any minimal/maximal length or minimal momentum.
However, one can wonder about  a simultaneous measurement of  $\hat X$ and $\hat P_y$? 
Let   say that, a simultaneous $\hat X,\hat P_y$-measurement is less straightforward  since 
terms of the type $\langle   \hat Y \hat X \rangle$ and   $\langle   \hat Y \hat P_y \rangle$    
are  encountered which cannot be treated in the same manner. Furthermore,
since the behaviour of $\hat X$ and $\hat P_y$ is linear on both sides of the inequality
in both cases, we do not expect a minimal/maximal length or a minimal momentum to arise in this circumstance.

\subsection{ Hilbert space representation with uncertainty relations }

As we mentioned in the previous subsection, the emergence of minimal 
  length $\Delta X_{min}^{abs}$ and minimal momentum $\Delta P_{y_{min}}^{abs}$ lead to the lost of
  representation of  the wavefunctions  in $X$ and $P_y$ directions  respectively,  except the 
 representation in $Y$ direction.  In the following, let studies  the representation of operators with uncertainty measurements.
 
 \subsubsection{ Representation with maximal length and minimal momentum}
 
 In the case  of the  uncertainty relation (\ref{in2}) that predicts a
  maximal  length and a minimal momentum, deduced from the relation
  $ [\hat Y,\hat P_y ]=i\hbar (1-\tau \hat Y +\tau^2 \hat Y^2) $  can
   be defined by the operators
 \begin{eqnarray}
  \hat Y&=&\hat y_0,\\
   \hat P_y&=&(\mathbb{I}-\tau \hat y_0+\tau^2\hat y_0^2)\hat p_{y_0},
 \end{eqnarray} 
 where $ \hat p_{y_0}=-i\hbar \partial_{y_0}$.  Then by operating on position space
wave function $ \psi(y_0)$, we have
 \begin{eqnarray}
  \hat Y \psi(y_0)&=& y_0 \star\psi(y_0),\\
   \hat P_y \psi(y_0)&=&-i\hbar(1-\tau  y_0+\tau^2 y_0^2)\partial_{y_0} \psi(y_0).
 \end{eqnarray} 
By utilizing the asymmetric Bopp-shift $\mathcal{B}_\theta^{a_1}$, these equations become
\begin{eqnarray}
  \hat Y \psi(y_s)&=& y_s \psi(y_s),\label{rep1}\\
   \hat P_y \psi(y_s)&=&-i\hbar(1-\tau  y_s+\tau^2 y_s^2)\partial_{y_s} \psi(y_s) \label{rep2},
 \end{eqnarray} 
 where  $  \psi(y_s)$ is defined on dense domain  $S_{\infty} $ of functions decaying faster
than any power.  Evidently, in  this deformed space, the position operator 
 is symmetric and self-adjoint while the momentum operator is not. Thus,
 the Hermiticity requirement of the momentum operator leads to the following proposition :\\
 
  {\bf Proposition 3.2.1.} {\it For the given completeness relation   on the complete  basis $\{|y_s\rangle\}$
  such as 
\begin{eqnarray} \label{id} 
 \int_{-l_max}^{l_max}\frac{dy_s}{(1-\tau  y_s+\tau^2 y_s^2)} |y_s\rangle\langle y_s|&=&\mathbb{I},
\end{eqnarray}
we have 
\begin{eqnarray}
  \langle \phi|\hat P_y\psi\rangle=  \langle \hat P_y^\dag \phi|\psi\rangle,
\end{eqnarray}}
such as 
\begin{eqnarray}
 \mathcal{D}(\hat P_y)&=& \left\{\psi,\psi'\in \mathcal{L}^2(-l_{max},l_{max}); \psi(-l_{max})= \psi'(l_{max})=0\right\},\\
  \mathcal{D} (\hat P_y^\dag )&=& \left\{\phi,\phi'\in \mathcal{L}^2(-l_{max},l_{max}) \right\}.
\end{eqnarray}\\\\
 {\it Proof.}  Let  consider $\psi \in \mathcal{D}(\hat P_y)$ and $ \phi \in \mathcal{D}(\hat P_y^\dag)$
 \begin{eqnarray}
  \langle \phi|\hat P_y\psi\rangle &=& \int_{-l_max}^{l_max}\frac{dy_s}{(1-\tau  y_s+\tau^2 y_s^2)} \phi^*(y_s)\left[
  -i\hbar(1-\tau  y_s+\tau^2 y_s^2)\partial_{y_s} \psi(y_s)\right].
  \end{eqnarray}
By  performing a partial integration, we have
  \begin{eqnarray}
    \langle \phi|\hat P_y\psi\rangle &=&  \int_{-l_max}^{l_max}\frac{dy_s}{(1-\tau  y_s+\tau^2 y_s^2)} 
    \left[
  -i\hbar(1-\tau  y_s+\tau^2 y_s^2)\partial_{y_s} \phi(y_s)\right]^*\psi(y_s)\cr &&
  +\left[-i\hbar \phi^*(y_s)\psi(y_s)\right]_{-l_max}^{l_max}\cr
   &=&  \langle \hat P_y^\dag \phi|\psi\rangle,
 \end{eqnarray}
where  $\psi(y_s)$ vanishes at $\pm l_{max}$ then $\phi^*(y_s)$ can attain any arbitrary value at the boundaries. 
The above equation implies that $\hat P_y$ is symmetric but it is not a self-adjoint operator. The situation
is that, $\hat P_y$ is a derivative operator on an interval with Dirichlet boundary conditions and all the 
candidates for the eigenfunctions of $\hat P_y$ are not in the domain of $\hat P_y$ because
they obey no longer the Dirichlet boundary conditions \cite{35}. In fact, the domain of $\hat P_y^\dag$ is much larger than that of 
$\hat P_y$, so $\hat P_y$  is indeed not self-adjoint.

Consequently, the scalar product between two states $|\Psi\rangle$ and $ | \Phi\rangle$  and the orthogonality of position eigenstate become
\begin{eqnarray}
 \langle \Phi|\Psi\rangle& =& \int_{-l_max}^{l_max}\frac{dy_s}{(1-\tau  y_s+\tau^2 y_s^2)}\Phi^*(y_s)\Psi(y_s),\\
 \langle y_s|y_s'\rangle &=& (1-\tau  y_s+\tau^2 y_s^2)\delta(y_s-y_s').
\end{eqnarray}
For $\tau\rightarrow  0$, we recover the usual completeness and orthogonality relations of bounded space
$ \mathcal{L}^2( -l_{max}, l_{max})$.

In order to give an explicite expression of the  eigenfunction $\psi(y_s)$, one 
solves the eigenvalue problem 
\begin{eqnarray}
   \hat P_y \psi_\zeta(y_s)=\zeta  \psi_\zeta (y_s).
 \end{eqnarray}
By solving the following differential equation
\begin{eqnarray}
-i\hbar(1-\tau  y_s+\tau^2 y_s^2)\frac{\partial \psi_\zeta(y_s)}{\partial y_s}= \zeta  \psi_\zeta (y_s),
\end{eqnarray}
we obtain the  position eigenvectors in the  form
\begin{eqnarray}\label{fzeta}
 \psi_\zeta (y_s)=  \psi_\zeta (0)\exp\left(i\frac{2\zeta}{\tau\hbar \sqrt{3}}\left[\arctan\left(\frac{2\tau y_s-1}{\sqrt{3}}\right)
 +\arctan\left(\frac{1}{\sqrt{3}}\right)\right]\right).
\end{eqnarray}
Then by normalization, $\langle \psi_\zeta|\psi_\zeta\rangle=1$, we have 
\begin{eqnarray}
 1&=&\int_{-l_max}^{l_max}\frac{dy_s}{(1-\tau  y_s+\tau^2 y_s^2)} \psi_\zeta^* (y_s) \psi_\zeta (y_s)\cr
  &=& |\psi_\zeta (0)|^2 \int_{-l_max}^{l_max}\frac{dy_s}{(1-\tau  y_s+\tau^2 y_s^2)}.
\end{eqnarray}
so, we find
\begin{eqnarray}\label{nzeta}
 \psi_\zeta (0)&=&\sqrt{\frac{\tau\sqrt{3}}{2}} \left[\arctan\left(\frac{2\tau l_{max}-1}{\sqrt{3}}\right)
 +\arctan\left(\frac{2\tau l_{max}+1}{\sqrt{3}}\right)\right]^{-\frac{1}{2}}\cr
 &=& \sqrt{\frac{\tau\sqrt{3}}{\pi}}.
\end{eqnarray}
Substituting this equation (\ref{nzeta}) into the equation (\ref{fzeta}), we have 
\begin{eqnarray}
 \psi_\zeta (y_s) &=& \sqrt{\frac{\tau\sqrt{3}}{\pi}} \exp\left(i\frac{2\zeta}{\tau\hbar \sqrt{3}}\left[\arctan\left(\frac{2\tau y_s-1}{\sqrt{3}}\right)
 +\arctan\left(\frac{1}{\sqrt{3}}\right)\right]\right).
\end{eqnarray}
This is the generalized position space eigenstate of
the position operator in the presence of both  minimal
momentum and  maximal length. In comparison with the seminal result of   Nozari and Etemadi \cite{26} done on momentum space, our result  slightly fits with theirs.

Let note that, the  goal of  this framework is to show,  how looks   the passing from the    position representation (\ref{fzeta})
to  the momentum representation. Therefore, the transformation that maps position space wave
functions into momentum space wave functions is  the Fourier transformation. The situation is that, 
the appearance of the minimal momentum given by Eq.(\ref{minmom}) leads to a loss of the notion of localized momentum
states  since we cannot probe the momentum space with a resolution less than the minimal momentum. So, 
 to treat this  problem in a realistic manner, we are forced to introduce the maximal momentum localization states
 that  let  information on momentum space  accessible.
 
 Now we consider    the maximal  localization states denoted by  $|\psi_\gamma^{max}\rangle$ 
  defined as states localized around a momentum $\gamma$, such that we have
  \begin{eqnarray}
   \langle \psi_\gamma^{max}|\hat P_y|\psi_\gamma^{max}\rangle= \gamma
  \end{eqnarray}
and are solutions of the following equation:
\begin{eqnarray}\label{diff}
\left(\hat P_y- \langle \hat P_y\rangle+\frac{\langle [\hat Y,\hat P_y]\rangle }{2\Delta Y^2 }\left(\hat Y-\langle \hat Y \rangle \right)\right)
|\psi_\gamma^{max}\rangle=0.
\end{eqnarray}
Using Eqs.(\ref{rep1}) and (\ref{rep2}), the differential equation in position space corresponding to (\ref{diff}) is in the following
form
\begin{eqnarray}
 \left(-i\hbar(1-\tau  y_s+\tau^2 y_s^2)\partial_{y_s}- 
 \langle \hat P_y\rangle+i\hbar\frac{1-\tau \langle \hat Y\rangle+\tau^2 \Delta Y^2+\tau^2 \langle \hat Y\rangle^2 }{2\Delta Y^2 }  
 (y_s  -\langle \hat Y \rangle)
 \right)\cr \times\psi_\gamma^{max}(y_s)=0.
\end{eqnarray}
 The solution to this equation is given by
  \begin{eqnarray}
  \psi_\gamma^{max}(y_s)=\Psi e^{  \frac{2}{\tau\hbar \sqrt{3}}\left[\frac{\hbar}{2\Delta Y^2} \left(\frac{1}{2\tau}-\langle \hat Y\rangle\right) 
  \left(1-\tau \langle \hat Y\rangle+\tau^2 \Delta Y^2+\tau^2 \langle \hat Y\rangle^2\right)+i\langle \hat P_y\rangle\right]
  \left(\arctan(\frac{2\tau y_s-1}{\sqrt{3}})
 +\arctan(\frac{1}{\sqrt{3}})\right)},
 \end{eqnarray}
 where 
 \begin{eqnarray}
  \Psi =  \psi_\gamma^{max}(0) (1-\tau  y_s+\tau^2 y_s^2)^{\frac{1-\tau \langle \hat Y\rangle+\tau^2 
  \Delta Y^2+\tau^2 \langle \hat Y\rangle^2 }{4\tau^2\Delta Y^2 } }.
 \end{eqnarray}
The states of absolutely maximal momentum localization are those with $\langle \hat P_y\rangle= \gamma$, 
 $\langle \hat Y\rangle= 0$ and if we restrict these states to the ones for which $\Delta Y=\frac{1}{\tau}$, we obtain
 \begin{eqnarray}
  \psi_\gamma^{max}(y_s) &=&   \psi_\gamma^{max}(0) (1-\tau  y_s+\tau^2 y_s^2)^{\frac{1}{2} }
  e^{  \frac{1}{ \sqrt{3}}   \left(\arctan(\frac{2\tau y_s-1}{\sqrt{3}})
 +\arctan(\frac{1}{\sqrt{3}})\right)}\cr&&\times
 e^{  i\frac{2\gamma}{ \tau\hbar\sqrt{3}}   \left(\arctan(\frac{2\tau y_s-1}{\sqrt{3}})
 +\arctan(\frac{1}{\sqrt{3}})\right)}.
 \end{eqnarray}
 To determine $ \psi_\gamma^{max}(0)$, we normalize to unity, 
  $\langle \psi_\gamma^{max}|\psi_\gamma^{max} \rangle=1$, we find
  \begin{eqnarray}
   1&=&\int_{-l_{max}}^{l_{max}}\frac{dy_s}{(1-\tau  y_s+\tau^2 y_s^2)} {\psi_\gamma^*}^{max}(y_s) \psi_\gamma^{max}(y_s)\cr
   &=&   {\psi_\gamma^*}^{max}(0) \psi_\gamma^{max}(0) \int_{-l_{max}}^{l_{max}}dy_s
   e^{  \frac{2}{ \sqrt{3}}   \left(\arctan(\frac{2\tau y_s-1}{\sqrt{3}})
 +\arctan(\frac{1}{\sqrt{3}})\right)},
  \end{eqnarray}
which gives
\begin{eqnarray}
  \psi_\gamma^{max}(0) &=&A^{-1/2}\cr&&\times
  \left[B  (3e^{\alpha_1}+
  e^{\alpha_2}) +C(e^{\alpha_2}\mathcal{F}^1-
  e^{\alpha_1}\mathcal{F}^2)+
  \sqrt{2}(e^{-i\frac{\pi}{3}-\alpha_1}\mathcal{F}^3-
  e^{i\frac{\pi}{3}+\alpha_2}\mathcal{F}^4)                \right]^{-1/2}, 
 \end{eqnarray}
 where
 \begin{eqnarray}
 A&=& \frac{\sqrt{3}}{2\tau(i\sqrt{2}-2)},\quad
 B= \frac{i}{\sqrt{3}(2i+\sqrt{2})},\quad
 C= (2i+\sqrt{2}),\\
 \alpha_1&=& -\frac{\pi\sqrt{2}}{3},\quad
 \alpha_2= \frac{\pi\sqrt{2}}{6},\,
 \mathcal{F}^1={}_{2}F_1(1,-\frac{i}{\sqrt{2}},1-\frac{i}{\sqrt{2}},-e^{i\frac{\pi}{3}}),\\
 \mathcal{F}^2&=&{}_{2}F_1(1,-\frac{i}{\sqrt{2}},1-\frac{i}{\sqrt{2}},-e^{i\frac{2\pi}{3}}),\quad
 \mathcal{F}^3={}_{2}F_1(1,-\frac{i}{\sqrt{2}},2-\frac{i}{\sqrt{2}},-e^{i\frac{\pi}{3}}),\\
 \mathcal{F}^4&=&{}_{2}F_1(1,-\frac{i}{\sqrt{2}},2-\frac{i}{\sqrt{2}},-e^{-i\frac{2\pi}{3}}).
  \end{eqnarray}

 Therefore,  the position space wave functions  of the states
that are maximally localized around a momentum $\gamma$  are in the following form
\begin{eqnarray}\label{max1}
  \psi_\gamma^{max}(y_s) &=& \psi_\gamma^{max}(0) \sqrt{1-\tau y_s+\tau^2 y_s^2}  e^{  \frac{1}{ \sqrt{3}}   \left(\arctan(\frac{2\tau y_s-1}{\sqrt{3}})
  	+\arctan(\frac{1}{\sqrt{3}})\right)}\cr&&\times
  e^{  i\frac{2\gamma}{ \tau\hbar\sqrt{3}}   \left(\arctan(\frac{2\tau y_s-1}{\sqrt{3}})
  	+\arctan(\frac{1}{\sqrt{3}})\right)}.
 \end{eqnarray}
 By projecting arbitrary  states onto this  maximally localized states (\ref{max1}) we recover information about the localization around the momentum. This 
 procedure is known as the concept of quasi representation wavefunction. We
take $|\chi\rangle$ as an arbitrary state, then the probability amplitude on maximal localization states around the momentum
$\gamma$ is  $\langle  \psi_\gamma^{max}|\chi\rangle=\chi(\gamma)$  namely quasi-momentum wavefunction.
 Thus, the passing  from the  position-space wave function into its
quasi representation wave function now would be

\begin{eqnarray}
 \chi(\gamma)&=& \psi_\gamma^{max}(0)\int_{-l_{max}}^{l_{max}}\frac{dy_s}{(1-\tau y_s+\tau^2 y_s^2)^\frac{1}{2}} e^{  \frac{1}{ \sqrt{3}}   \left(\arctan(\frac{2\tau y_s-1}{\sqrt{3}})
 	+\arctan(\frac{1}{\sqrt{3}})\right)}\cr&&\times
 e^{  -i\frac{2\gamma}{ \tau\hbar\sqrt{3}}   \left(\arctan(\frac{2\tau y_s-1}{\sqrt{3}})
 	+\arctan(\frac{1}{\sqrt{3}})\right)}  \chi(y_s).
\end{eqnarray}
This transformation that maps position space wave
functions into quasi-momentum space wave functions is the
generalization of the Fourier transformation. The inverse transformation is given by
\begin{eqnarray}
 \chi(y_s)&=&\int_{-\infty}^{\infty}d\gamma\frac{[2\pi \hbar \psi_\gamma^{max}(0)]^{-1}}{(1-\tau y_s+\tau^2 y_s^2)^{\frac{1}{2}}} e^{ - \frac{1}{ \sqrt{3}}   \left(\arctan(\frac{2\tau y_s-1}{\sqrt{3}})
 	+\arctan(\frac{1}{\sqrt{3}})\right)}\cr&&\times
 e^{  i\frac{2\gamma}{ \tau\hbar\sqrt{3}}   \left(\arctan(\frac{2\tau y_s-1}{\sqrt{3}})
 	+\arctan(\frac{1}{\sqrt{3}})\right)} \chi(\gamma).
\end{eqnarray}

\subsubsection{Representation with maximal and minimal lengths}
\subsubsection*{ $\diamond$ {\it Representation on position space}}
  From the relation 
  $ [\hat X,\hat Y ]=i\hbar (1-\tau \hat Y +\tau^2 \hat Y^2) $ that predicts 
  maximal and minimal lengths  can
   be defined by the operators
   \begin{eqnarray}
  \hat Y&=&\hat y_0,\\
   \hat X&=&(\mathbb{I}-\tau \hat y_0+\tau^2\hat y_0^2)\hat x_0.
 \end{eqnarray} 
Using again the asymmetric Bopp-shift $\mathcal{B}_\theta^{a_1}$ and acting these operators one the wave function $\psi(y_s)$, we have 
\begin{eqnarray}
  \hat Y \psi(y_s)&=& y_s \phi(y_s),\label{rep3}\\
   \hat X \psi(y_s)&=&\left(1-\tau  y_s+\tau^2 y_s^2\right) x_s\phi(y_s)+
   \frac{i\theta}{2} \left(1-\tau  y_s+\tau^2 y_s^2\right)\partial_{y_s} \phi(y_s)\label{rep4}.
 \end{eqnarray} 
 Based one the equation (\ref{id}),  one can state the following proposition:\\\\
  {\bf Proposition 3.2.1.} {\it  The operator  $\hat X$   on the dense domain $\mathcal{D}(\hat X)$ is symmetric such as
\begin{eqnarray}
 \langle \psi|\hat X\phi\rangle =  \langle  \hat X^\dag \psi|\phi\rangle,
\end{eqnarray}
but is not self-adjoint  
\begin{eqnarray}
 \mathcal{D}(\hat X)&=& \left\{\phi,\phi'\in \mathcal{L}^2(-l_{max},l_{max}); \phi(-l_{max})= \phi'(l_{max})=0\right\},\\
  \mathcal{D} (\hat X^\dag )&=& \left\{\psi,\psi'\in \mathcal{L}^2(-l_{max},l_{max}) \right\}.
\end{eqnarray}}

\subsubsection*{ $\diamond$ {\it Position eigenfunction}}
The position operator $\hat X$ acting on the operator $\hat Y$
eigenstates gives
\begin{eqnarray}
   \hat X \phi_\lambda (y_s)=\lambda  \phi_\lambda (y_s).
 \end{eqnarray}
By solving the following differential equation  
  \begin{eqnarray}
   \frac{i\theta}{2} \left(1-\tau  y_s+\tau^2 y_s^2\right)\partial_{y_s} \phi_\lambda (y_s)=
   \left[\lambda-\left(1-\tau  y_s+\tau^2 y_s^2\right) x_s\right]\phi_\lambda (y_s),
 \end{eqnarray}
 we obtain
 \begin{eqnarray}
 \phi_\lambda (y_s)=  \phi_\lambda (0)\exp\left(-i\frac{4\lambda}{\tau\theta \sqrt{3}}\left[\arctan\left(\frac{2\tau y_s-1}{\sqrt{3}}\right)
 +\arctan\left(\frac{1}{\sqrt{3}}\right)\right]+i\frac{2x_s}{\theta}y_s\right).
\end{eqnarray}
 Through the normalization of this function, 
we have
\begin{eqnarray}
 \phi_\lambda (y_s)&=&  \sqrt{\frac{\tau\sqrt{3}}{\pi}}  e^{-i\left(\frac{4\lambda}{\tau\theta \sqrt{3}}\left[\arctan\left(\frac{2\tau y_s-1}{\sqrt{3}}\right)
 +\arctan\left(\frac{1}{\sqrt{3}}\right)\right]-\frac{2x_s}{\theta}y_s\right)}.
\end{eqnarray}

\subsubsection*{ $\diamond$ {\it Maximal localization}}
Now we consider  $|\phi_\eta^{max}\rangle$ the states  of maximal localization around a position $\eta$ such as
\begin{eqnarray}
   \langle \phi_\eta^{max}|\hat X|\psi_\eta^{max}\rangle= \eta, 
  \end{eqnarray}
and are solution of the equation
\begin{eqnarray}\label{difff}
  \left( \hat X- \langle \hat X\rangle  +\frac{\langle [\hat X,\hat Y]\rangle }{2\Delta Y^2 }
 \left(\hat Y-\langle \hat Y \rangle \right)
 \right) |\phi_\gamma^{max}\rangle= 0.
\end{eqnarray}
Using Eqs.(\ref{rep3}) and (\ref{rep4}), the differential equation in position space corresponding to (\ref{difff}) is in the following
form
\begin{eqnarray}
 \left(1-\tau  y_s+\tau^2 y_s^2\right) x_s\phi_\eta^{max}(y_s)\cr+
   \left(\frac{i\theta}{2} \left(1-\tau  y_s+\tau^2 y_s^2\right)\partial_{y_s}  -\langle \hat X\rangle      
   +i\theta\frac{1-\tau \langle \hat Y \rangle +\tau^2\langle \hat Y \rangle ^2+\tau^2 \Delta Y^2 }{2\Delta Y^2 }(y_s
   -\langle \hat Y\rangle )\right)\cr \times\phi_\eta^{max}(y_s)&=&0.\cr
\end{eqnarray}
We obtain the states of maximal localization as follows
\begin{eqnarray}
 \phi_\eta^{max}=\Phi e^{i\frac{2x_s}{\theta}y_s}e^{-\frac{4}{\theta \tau\sqrt{3}} \left[\frac{\theta}{2\Delta Y^2}
 \left(\frac{1}{2\tau}-\langle \hat Y \rangle\right)\left(1-\tau \langle \hat Y \rangle +\tau^2\langle \hat Y \rangle ^2+\tau^2 \Delta Y^2 \right) +
 i \langle \hat X \rangle             
 \right]\left( \arctan\left(\frac{2\tau y_s-1}{\sqrt{3}}\right)
 +\arctan\left(\frac{1}{\sqrt{3}}\right)\right)},
\end{eqnarray}
where
\begin{eqnarray}
 \Phi=\phi_\eta^{max}(0) \left(1-\tau  y_s+\tau^2 y_s^2\right)^
 {-\frac{1-\tau \langle \hat Y \rangle +\tau^2\langle \hat Y \rangle ^2+\tau^2 \Delta Y^2 }{2\tau^2 \Delta Y^2}}.
\end{eqnarray}
The states of absolutely maximal localization are those with $\langle \hat X\rangle= \eta$, 
 $\langle \hat Y\rangle= 0$ and if we restrict these states to the ones for which $\Delta Y=\frac{1}{\tau}$, we obtain
 \begin{eqnarray}
  \phi_\eta^{max}&=& \phi_\eta^{max}(0) \left(1-\tau  y_s+\tau^2 y_s^2\right)^{-1}\ e^{i\frac{2x_s}{\theta}y_s}
  e^{-\frac{2}{\sqrt{3}} \left( \arctan\left(\frac{2\tau y_s-1}{\sqrt{3}}\right)
 +\arctan\left(\frac{1}{\sqrt{3}}\right)\right)}\cr&&\times
 e^{-i\frac{4\eta }{\tau \theta \sqrt{3}} \left( \arctan\left(\frac{2\tau y_s-1}{\sqrt{3}}\right)
 +\arctan\left(\frac{1}{\sqrt{3}}\right)\right)}.
 \end{eqnarray}
  By  normalization to unity, 
  $\langle \phi_\eta^{max}|\phi_\eta^{max} \rangle=1$, we find
  ¨
 \begin{eqnarray}
  \phi_\eta^{max}(0)=\left( \frac{43e^{\frac{4\pi}{3\sqrt{3}}}}{126\tau}  -
  \frac{3e^{\frac{-2\pi}{3\sqrt{3}}}}{14\tau}   \right)^{-\frac{1}{2}}.
  \end{eqnarray}
Therefore,  the position space wave functions  of the states
that are maximally localized around a momentum $\eta$  are in the following form
\begin{eqnarray}\label{max1}
  \phi_\eta^{max}(y_s) &=&\frac{ \phi_\eta^{max}(0)}{1-\tau y_s+\tau^2 y_s^2} \ e^{i\frac{2x_s}{\theta}y_s}
  e^{-\frac{2}{\sqrt{3}} \left( \arctan\left(\frac{2\tau y_s-1}{\sqrt{3}}\right)
  	+\arctan\left(\frac{1}{\sqrt{3}}\right)\right)}\cr&&\times
  e^{-i\frac{4\eta }{\tau \theta \sqrt{3}} \left( \arctan\left(\frac{2\tau y_s-1}{\sqrt{3}}\right)
  	+\arctan\left(\frac{1}{\sqrt{3}}\right)\right)}.
 \end{eqnarray}

\subsubsection*{ $\diamond$ {\it The generalization of the Fourier transformation and its inverse
}}
The generalized    Fourier transform obtained from 
  the passing  of  the  position-space wave function into 
quasi representation wave function $\langle  \phi_\eta^{max}|\rho\rangle=\rho(\eta)$ 
 is given by 

\begin{eqnarray}
 \rho(\eta)&=& \phi_\eta^{max}(0)\int_{-l_{max}}^{l_{max}}\frac{dy_s}{(1-\tau y_s+\tau^2 y_s^2)^2} \ e^{-i\frac{2x_s}{\theta}y_s}
 e^{-\frac{2}{\sqrt{3}} \left( \arctan\left(\frac{2\tau y_s-1}{\sqrt{3}}\right)
 	+\arctan\left(\frac{1}{\sqrt{3}}\right)\right)}\cr&&\times
 e^{i\frac{4\eta }{\tau \theta \sqrt{3}} \left( \arctan\left(\frac{2\tau y_s-1}{\sqrt{3}}\right)
 	+\arctan\left(\frac{1}{\sqrt{3}}\right)\right)}  \rho(y_s).
\end{eqnarray}
 and the inverse transformation is given by
 \begin{eqnarray}
 \rho(y_s)&=& 
 \int_{-\infty}^{+\infty}d\eta \frac{1-\tau y_s+\tau^2 y_s^2}{\pi\theta \phi_\eta^{max}(0)}   e^{i\frac{2x_s}{\theta}y_s}
 e^{\frac{2}{\sqrt{3}} \left( \arctan\left(\frac{2\tau y_s-1}{\sqrt{3}}\right)
 	+\arctan\left(\frac{1}{\sqrt{3}}\right)\right)}\cr&&\times
 e^{-i\frac{4\eta }{\tau \theta \sqrt{3}} \left( \arctan\left(\frac{2\tau y_s-1}{\sqrt{3}}\right)
 	+\arctan\left(\frac{1}{\sqrt{3}}\right)\right)}  \rho(\eta).
 \end{eqnarray}

 \subsection{ Decoupled and reduction into commutative space  }

Another possibility of representation of wave functions  is to decouple directly  the set of operators  $\hat X,\hat Y, \hat P_x,\hat P_y\ $  
in terms of operators  $\hat x_s,\hat y_s, \hat p_{x_s},\hat p_{y_s}$ using the transformations 
$\mathcal{R}_\tau$ and $\mathcal{B}_\theta^s$. 
 We  find
 \begin{eqnarray}
 \hat X&=&\hat x_s-\frac{\theta}{2\hbar}\hat p_{y_s}-\tau\hat y_s\hat x_s+\frac{\tau\theta}{2\hbar}(\hat y_s\hat p_{y_s}-\hat p_{y_s}\hat x_s)+
 \frac{\tau\theta^2}{4\hbar^2}\hat p_{y_s} \hat p_{x_s}
+ \tau^2\hat y_s^2\hat x_s \cr&&+\frac{\theta\tau^2}{2\hbar}\left(2\hat y_s\hat p_{x_s}\hat x_s-\hat y_s^2 \hat p_{y_s}\right)
+\frac{\theta^2\tau^2}{4\hbar^2}\left(\hat p_{x_s}^2\hat x_s-2\hat y_s\hat  p_{x_s}\hat  p_{y_s}\right)\cr&&-\frac{\theta^3\tau^2}{8\hbar^3} p_{x_s}^2 p_{y_s},\\
 \hat Y&=&\hat y_s+\frac{\theta}{2\hbar}\hat p_{x_s},\\
 \hat P_x&=&\hat p_{x_s},\\
 \hat P_y&=&\hat p_{y_s}-\tau\hat y_s\hat p_{y_s}-\frac{\tau\theta}{2\hbar}\hat p_{y_s}\hat p_{x_s}+
 \frac{\tau^2\theta}{\hbar}\hat y_s\hat p_{x_s}\hat p_{y_s}+\tau^2\hat y_s^2\hat p_{y_s}+\frac{\tau^2\theta^2}{4\hbar^2}\hat p_{x_s}^2\hat p_{y_s}.
\end{eqnarray}
From these representations,  follows immediately that  the operators  $\hat X$ and $\hat P_y$    are no
longer Hermitian in the space in which the operators $\hat x_s,\hat y_s, \hat p_{x_s},\hat p_{y_s}$ are Hermitian.
An immediate consequence is that Hamiltonian of  models formulated in terms of these operators  will in
general also not be Hermitian. In order to map these non Hermitian operators  $\hat X$ and $\hat P_y$ into Hermitian ones, 
%
%
we proceed by approximations in a first order of parameters $\theta$ and $\tau$ that we assume   very small. Therefore we obtain through the approximations of these operators an effective noncommutative space which is 
connected to $\kappa$-like realisations and  to the deformed Heisenberg algebra \cite{27,28,29,30,31}
\begin{eqnarray}\label{re4}
 \hat X&=&\hat x_s-\frac{\theta}{2\hbar}\hat p_{y_s}-\tau\hat y_s\hat x_s,\quad
 \hat Y=\hat y_s+\frac{\theta}{2\hbar}\hat p_{x_s},\cr
 \hat P_x&=&\hat p_{x_s},\quad\quad
 \hat P_y=\hat p_{y_s}-\tau\hat y_s\hat p_{y_s}.
\end{eqnarray}
It is easy to verify that these   operators   are Hermitian except  the operator $\hat P_y$ that one needs to symmetrize in order to guarantee 
the complete Hermiticity of this space.\\ 

 {\bf Proposition 3.3.} {\it For the  given completeness  relation 
\begin{eqnarray}  
 \int_{-\infty}^{+\infty}\frac{dx_sdy_s}{(1-\tau  y_s)} |x_s,y_s\rangle\langle x_s, y_s|&=&\mathbb{I},
\end{eqnarray}
with $|x_s,y_s\rangle$ elements of the domain of $\hat P_y$ maximally dense in $\mathcal{L}^2(\mathbb{R}^2)$, we  have 
\begin{equation}
  \hat P_y=\hat P_y^\dag.
\end{equation}}

 From the actions of operators (\ref{re4}) on the wave function $\psi(x_s, y_s)$, 
 we can thus obtain the following  differential representations
 \begin{eqnarray}
  \hat X \psi(x_s, y_s)&=& (x_s+i\theta/2\partial_{y_s}-\tau y_sx_s)\psi(x_s, y_s),\\
  \hat Y \psi(x_s, y_s)&=& (y_s-i\theta/2\partial_{x_s})\psi(x_s, y_s),\\
  \hat P_x \psi(x_s, y_s)&=& -i\hbar \partial_{x_s}\psi(x_s, y_s),\\
  \hat P_y \psi(x_s, y_s)&=& -i\hbar\left(1-\tau y_s\right)\partial_{y_s}\psi(x_s, y_s),
 \end{eqnarray}
  and the corresponding  maximal domains
  \begin{eqnarray}
   \mathcal{D}(\hat X)&=&\{\psi(x_s, y_s)\in  \mathcal{L}^2(\mathbb{R}^2):  (x_s+i\theta/2\partial_{y_s}-\tau y_sx_s)\psi(x_s, y_s) \in \mathcal{L}^2(\mathbb{R}^2)                                   \},\\
     \mathcal{D}(\hat Y)&=&\{\psi (x_s, y_s)\in  \mathcal{L}^2(\mathbb{R}^2):  (y_s-i\theta/2\partial_{x_s})\psi(x_s, y_s)  \in  \mathcal{L}^2(\mathbb{R}^2) \},\\
       \mathcal{D}(\hat P_x)&=&\{\psi (x_s, y_s)\in  \mathcal{L}^2(\mathbb{R}^2):  -i\hbar \partial_{x_s}\psi(x_s, y_s)  \in  \mathcal{L}^2(\mathbb{R}^2)                    \},\\
         \mathcal{D}(\hat P_y)&=&\{\psi (x_s, y_s)\in  \mathcal{L}^2(\mathbb{R}^2):    -i\hbar\left(1-\tau y_s\right)\partial_{y_s}\psi(x_s, y_s) 
         \in  \mathcal{L}^2(\mathbb{R}^2 )                \}.
  \end{eqnarray}
From  the solutions of the above  differential  equations, 
one can straightfowardly deduce  the corresponding   Fourier transforms. We leave this part  to the reader to determine these  transformation basing on the formulae (\ref{Four}).


  Notice that the set of   deformed operators (\ref{re4})  is less restrictive  than the representation (\ref{reR}) because  the latter leads to 
the minimal uncertainty measurements while the representation (\ref{re4}) does not present any ambiguity in the meaning of wavefunction.
It now depends on our choice   to treat  models in  the representation of preference.
In what  follows,  we use the representation (\ref{re4}) to  illustrate   the study of  some simple models in quantum mechanics. 

\section{Models in position dependent noncommutative space} \label{sec4}

The models of interest are the free particle, the particle in a box and  the harmonic oscillator. We  start by formulating them
in terms of operators $\hat X,\hat Y, \hat P_x,\hat P_y\ $ and then 
determine  how to solve the Schr\"{o}dinger equation exactly or  pertubately. 
Now, let consider $\hat H$ the Hamiltonian of a system of mass $m$  defined as follows

\begin{eqnarray}\label{Ham1}
 \hat H(\hat P_x,\hat P_y, \hat X,\hat Y):= \frac{1}{2m}(\hat P_x^2+\hat P_y^2)+ V(\hat X,\hat Y),
 \end{eqnarray}
  where $V$ is the potential energy of the system.  Using the relations (\ref{re4}), this Hamiltonian is
decoupled in terms of the following Hamiltonians
\begin{eqnarray}\label{Ham2}
 \hat H=\hat H_s+ \hat H_\theta +\hat H_\tau 
\end{eqnarray}
where $\hat H_s$ is the non-pertubated Hamiltonian, $\hat H_\tau$ and $\hat H_\theta$ are respectively  the  $\tau$-perturbation and 
  $\theta$-perturbation Hamiltonians. Let  stress that the Hamiltonians (\ref{Ham1}) and (\ref{Ham2}) are just different points
  of view to describe the same type of physics and in what follows, we will use the form (\ref{Ham2}) to solve the eigenvalue problems. 
\subsection{The free particle}
 The free particle Hamiltonian reads
 \begin{equation}
  \hat H(\hat X,\hat Y, \hat P_x,\hat P_y)= \frac{1}{2m}(\hat P_x^2+\hat P_y^2).
 \end{equation}
In the form (\ref{Ham2}), this Hamiltonian reads as
\begin{eqnarray}\label{Ham3}
 \hat H (\hat x_s,\hat y_s, \hat p_{x_s},\hat p_{y_s}) &=& \frac{1}{2m}\hat p_{x_s}^2+\frac{1}{2m}\hat p_{y_s}^2-\frac{\tau}{2m}\left[2y_s p_{y_s}^2-i\hbar p_{y_s}\right]\cr&&+
 \frac{\tau^2}{2m}\left[y_s^2 p_{y_s}^2-i\hbar y_s p_{y_s}\right].
\end{eqnarray}
 The  Schr\"{o}dinger equation is given by
 \begin{equation}\label{schr}
   \hat H \psi(x_s,y_s)=E \psi(x_s,y_s).
 \end{equation}
As it is clearly seen, the system is decoupled and the solution to the eigenvalue equation (\ref{schr})
is given by
\begin{eqnarray}
 \psi(x_s,y_s)= \psi_k(x_s)\psi_n(y_s),\quad  E= E_k+E_n
\end{eqnarray}
where $\psi_k(x_s)$ is the wave function in the $x_s$-direction and $\psi_n(y_s)$ the wave function in the $y_s$-direction. 
Since the particle is free in the $x_s$-direction, the wave function is \cite{23}
\begin{eqnarray}
 \psi_k(x_s)=\int_{-\infty}^{+\infty} dk g(k)e^{ikx_s},
\end{eqnarray}
where $g(k)$ determines the shape of the wave packet and the energy spectrum is continuous \cite{1,23}
\begin{eqnarray}
 E_k=\frac{\hbar^2 k^2}{2m}.
\end{eqnarray}
In $y_s$-direction, we have to solve  the following equation 
\begin{eqnarray}
 (1-\tau y_s)^2 \frac{d^2 \psi_n}{dy_s^2}-\tau(1-\tau y_s)  \frac{d \psi_n}{dy_s}+\frac{2m }{\hbar^2}
 E_n \psi_n=0.
\end{eqnarray}
By setting $(1-\tau y_s)=e^z$, the above equation is reduced into
\begin{eqnarray}
 \frac{d^2 \psi_n}{dz^2} +\lambda^2 \psi_n=0.
\end{eqnarray}
This equation is the equation of free harmonic oscillations with  $\lambda^2=\frac{2m }{\tau^2\hbar^2} E_n$ the frequency of  oscillation. The solution is given by
\begin{eqnarray}
\psi_n(y_s)&=& A\sin(\lambda z)+ B\sin(\lambda z)\cr
      &=& A\sin\left[\lambda \ln (1-\tau y_s)\right]+ B\cos\left[\lambda \ln (1-\tau y_s)\right],
\end{eqnarray}  
where $A$, $B$ are constantes and $\tau$ is considered very smaller than one. If we assume that, the frequency of  oscillation is quantized such  as $\lambda=2\pi n$ with 
$n\in\mathbb{N}^*$, therefore the engenvalue $E_n$ is given by
\begin{eqnarray}
E_n=\frac{2\pi^2\tau^2\hbar^2}{m}n^2.
\end{eqnarray}

\subsection{Particle in a box}
We  consider the above   free  particle of mass $m$   captured in a two-dimensional box of length $a$ and height $b$. The  boundaries of the box are
located at $ 0\leq x_s\leq a$ and $ 0\leq y_s\leq b$.
The above  Hamiltonian (\ref{Ham3}) is rewritten as follows 
 \begin{eqnarray}
  \hat H=
   \begin{cases}
    \hat H_s= \frac{1}{2m}(\hat p_{x_s}^2+\hat p_{y_s}^2),\\
    \hat H_\tau = -\frac{\tau}{2m}\left[2y_s p_{y_s}^2-i\hbar p_{y_s}\right]+
 \frac{\tau^2}{2m}\left[y_s^2 p_{y_s}^2-i\hbar y_s p_{y_s}\right].
     \end{cases}
  \end{eqnarray}
To  solve the   eigenvalue equation,   we may resort to the  perturbation theory to obtain 
some useful insight on the solutions. Thus, the eigenvalues and eigenfunctions of $\hat H_s$  are given by \cite{34}
\begin{eqnarray}
 E_s&=& \frac{\hbar^2\pi^2}{2m}\left[\frac{n_{x_s}^2}{a^2}+\frac{n_{y_s}^2}{b^2}\right],\\
 \psi_s(x_s,y_s)&=& \frac{2}{\sqrt{ab}}\sin\left( \frac{n_{x_s}\pi x_s}{a}\right)\sin\left( \frac{n_{y_s}\pi y_s}{b}\right),
\end{eqnarray}
$n_{x_s},n_{y_s}\in \mathbb{N}^*$ and $ab$ is just the area of the box. The wave functions 
  satisfy the Dirichlet  condition i.e it 
vanishes at the boundaries  $\psi_s(0)=\psi(a)=0$ and $\psi_s(0)=\psi_s(b)=0$.

Now,  for the sake of simplicity we restrict the Hamiltonian $\hat H_\tau$  to first order of the parameter $\tau$ which is given by
\begin{eqnarray}
 \hat H_\tau = -\frac{\tau}{2m}\left(2y_s p_{y_s}^2-i\hbar p_{y_s}\right) +\mathcal{O}(\tau).
\end{eqnarray}
Using the perturbation theory, we determine the effect $  E_\tau$ on the energy eigenvalues
\begin{eqnarray}
  E_\tau&=&  \langle \psi_s|\hat H_\tau|\psi_s\rangle
 =\frac{\tau\hbar^2}{2m}\int_0^a \int_0^b  \psi_s^*(x,y)\left( 2y_s\partial_{y_s}^2+\partial_{y_s} \right)\psi_s(x,y)dx_sdy_s\cr
              &=&    -\tau  \frac{\hbar^2\pi^2 n_y^2}{2mb}.
\end{eqnarray}
Comparing the $\tau$-corrections to the unperturbed
energy term  in the case where $a=b=L$ and  $n_{x_s}=n_{y_s}=n$, we get 
\begin{eqnarray}
\frac{| E_\tau|}{E_s}=\tau\frac{L}{2}.
\end{eqnarray}

\subsection{The harmonic oscillator}

The Hamiltonian of a two dimensional harmonic oscillator  is given by
\begin{eqnarray}
\hat H= \frac{1}{2m}(\hat P_x^2+\hat P_y^2)+\frac{1}{2}m\omega^2(\hat X^2+\hat Y^2).
\end{eqnarray}
Using the representation (\ref{re4}), the corresponding Hamiltonian
reads
\begin{eqnarray}
  \hat H=
   \begin{cases}
    \hat H_s= \frac{1}{2m}(\hat p_{x_s}^2+\hat p_{y_s}^2)+\frac{m\omega^2}{2}(\hat x_s^2+\hat y_s^2)\\
    \hat H_\tau = -\frac{\tau}{2m}\left(2\hat y_s \hat p_{y_s}^2-i\hbar \hat  p_{y_s}+2m^2\omega^2\hat y_s\hat x_s^2\right)\\
    \hat H_\theta= -\frac{m\omega^2\theta}{2\hbar}\hat L_z\\
    \hat H_{\tau^2}=\frac{\tau^2 m\omega^2}{2}\hat x_s^2\hat y_s^2\\
    \hat H_{\theta^2}=\frac{m\omega^2\theta^2}{8\hbar^2} (\hat p_{x_s}^2+\hat p_{y_s}^2)\\
     \hat H_{\tau\theta}=\frac{m\omega^2\tau\theta}{2\hbar}(2\hat y_s \hat p_{x_s}-i\hbar)\hat x_s
     \end{cases}
  \end{eqnarray}
 where $\hat L_z=\left(\hat x_s \hat p_{y_s}-\hat y_s \hat p_{x_s}\right)$ is the angular momentum. 
 It is important to remark that the $\theta$-pertubation introduced a 
 dynamical   $SO(2)$ rotations in the plan. Since $[\hat H_s, \hat H_\theta]=0$, to determine the corresponding  basis  
  which can diagonalize simultaneously these operators,  we consider the  helicity Fock algebra generators as follows
  \begin{eqnarray}
  a_\pm&=&\frac{m\omega}{2\hbar\sqrt{2}}\left[(\hat x_s\pm i \hat y_s)+\frac{i}{m\omega}\left(\hat p_{x_s}\pm i\hat p_{y_x}\right) \right],\\
  a_\pm^\dag&=&\frac{m\omega}{2\hbar\sqrt{2}}\left[(\hat x_s\pm i \hat y_s)-\frac{i}{m\omega}\left(\hat p_{x_s}\mp i\hat p_{y_x}\right) \right],
  \end{eqnarray}
  which satisfy 
  \begin{eqnarray}
 [a_\pm,a_\pm^\dag]=\mathbb{I},\quad  [a_\pm,a_\mp^\dag]=0.
\end{eqnarray}
The associated  orthonormalized   helicity  basis $|\psi_{n_+,n_-} \rangle$ are defined as follows
\begin{eqnarray}
|\psi_{n_+,n_-}\rangle&=&\frac{1}{\sqrt{n_-!n_+!}}\left(a_+^\dag\right)^{n_+}\left(a_-^\dag\right)^{n_-}|\psi_{0,0}\rangle\quad \mbox{and}\\
 \langle \psi_{m_+,m_-} |\psi_{n_+,n_-}\rangle&=&\delta_{m_+n_+}\delta_{m_-n_-},\quad 
 \sum_{n_\pm=0}^{+\infty}|\psi_{n_+,n_-}\rangle  \langle \psi_{n_+,n_-} |=\mathbb{I}.
\end{eqnarray}
The action of these operators reads as
\begin{eqnarray}
a_\pm|\psi_{n_\pm}\rangle&=&\sqrt{n_\pm}|\psi_{n_{\pm} - 1}\rangle,\\
a_{\pm}^\dag |\psi_{n_\pm}\rangle&=&\sqrt{n_\pm +1}|\psi_{n_\pm}+1\rangle,\\
a_{\pm}^\dag  a_\pm|\psi_{n_\pm}\rangle &=&n_\pm  |\psi_{n_{\pm}}\rangle.
\end{eqnarray}

Conversely, we have 
\begin{eqnarray}
\hat x_s&=&\frac{1}{2} \sqrt{\frac{\hbar}{m\omega}}\left[a_+ +a_-+a_+^\dag+a_-^\dag\right],\,
\hat y_s=\frac{i}{2} \sqrt{\frac{\hbar}{m\omega}}\left[a_+ -a_--a_+^\dag+a_-^\dag\right],\\
\hat p_{x_s}&=&-i\frac{m\omega}{2}\sqrt{\frac{\hbar}{m\omega}}\left[a_+ +a_--a_+^\dag-a_-^\dag\right],\cr&&
\hat P_{y_s}=\frac{m\omega}{2}\sqrt{\frac{\hbar}{m\omega}}\left[a_+ -a_-+a_+^\dag-a_-^\dag\right].
\end{eqnarray}
At first order of the parameters $\theta$ and $\tau$, the Hamiltonian is reduced into
\begin{eqnarray}
\hat H= \hat H_s+ \hat H_\theta+\hat H_\tau +\mathcal{O}(\tau)+ \mathcal{O}(\theta)
\end{eqnarray}
The energy eigenvalues for the Hamiltonian $\hat H_s$ and for the pertubated Hamiltonian
$\hat H_\theta$ and $\hat H_\tau$ reads as follows
\begin{eqnarray}
E_s&=& \hbar \omega \left(n_++n_-+1\right),\quad 
E_\theta=\frac{m\omega^2\theta}{2\hbar}\left(n_--n_+\right),\quad
E_\tau=0.
\end{eqnarray}
These results show that, for the case $E_\tau=0$, there is  no contribution in 
$\tau$-deformed  energy spectrum.   To improve this result we look at the second order in $\tau$-perturbation, namely
\begin{eqnarray}
 E_{\tau^2}= \sum_{k_\pm\neq n_\pm}^\infty \frac{\langle \psi_{n_\pm}| \hat H_\tau|\psi_{k_\pm}\rangle 
 \langle \psi_{k_\pm}|\hat H_\tau|\psi_{n_\pm}\rangle }{E_{n_\pm}^0-E_{k_\pm}^0}.
\end{eqnarray}
For the sake of simplicity,  this energy at  the ground states $n_\pm=0$ is evaluated at
\begin{eqnarray}
  E_{\tau^2}&=& \frac{\tau^2}{4m^2}\left(\frac{5 m\hbar^2}{12}+0+\frac{17m\hbar^2}{48}\right)\cr
            &=& \frac{37\hbar}{384m}\tau^2.
\end{eqnarray}

\section{Conclusion Remarks}\label{section5}
We have introduced a new version of position dependent noncommutative space-time in two dimensional configuration spaces. 
This    space-time that we provided, 
generalizes the set of noncommutative space-time recently introduced by   Fring {\it et al} \cite{1}. 
To construct this noncommutative space-time (\ref{alg3}),
we have considered  the most used deformed   commutative space-time (\ref{alg2})  
in such a way that  at the limit $\tau \rightarrow 0$ we recovered this   algebra (\ref{alg2}).  
The interesting physical consequence we found is that, this noncommutative space-time leads 
to     minimal and maximal lengths  for simultaneous measurement in $X,Y$-directions. Then for a 
simultaneous measurement in $Y,P_y$-directions, this space also leads  to a minimal momentum and a maximal length.
The existence of this  maximal length,  which is the basic difference to the work of  Fring {\it et al},  
is related to the presence of  an extra, first order term in particle’s length. It brings a lot of new features in the 
 representation of this noncommutation space. Moreover, to escape the difficulties from dealing with this  representation due to the presence
 of the minimal uncertainties, we propose another   representation of operators obtained by approximations in first order of  parameters $\theta$ and  $\tau$. In this  new  representation,
 we   provided the spectra  of  some fundamental quantum systems  such as  the free particle, 
the particle in a box  and the Harmonic oscillator.

 It is well known  that the presence of both minimal length and minimal momentum   raised the question of  singularity 
 of the space-time i.e the space is inevitably bounded by minimal quantities beyond which any further localization of particle 
 is not possible \cite{4}. With Fring \textit{et al}. noncommutative space-time, it  is  shown  that any  object in this  space will be string like i.e a   measurement in  $\hat X$ and $\hat Y$ spaces leads to a minimal length for $\hat X$ or for $\hat Y$ when  informations are given-up in one direction. In comparison  with  this  work, my version of noncommutative space-time 
introduces a singularity in $X$-direction  and a  broken   singularity in $\hat Y$-direction  for simultaneous measurement in both directions. 
This means that, the lost of localization of particle in $X$-direction  can be maximally recorved in   $Y$-direction. Furthermore the singularity 
in momentum  $P_y$-direction leads to the maximal localization in $\hat Y$-direction for a simultaneous measurement in both directions.

Moreover, looking at the representation $\mathcal{R}_\tau$ which generates the algebra (\ref{alg3}),  follows immediately that 
some  operators      are no  longer Hermitian in the space in which the operators $\hat x_0,\hat y_0, \hat p_{x_0},\hat p_{y_0}$ are Hermitian.  
 In order  to use the approximation method to map  these non Hermitian operators  into Hermitian ones in the space of standard Heisenberg operators,  we  may try to find a similarity transformation, i.e. a Dyson map \cite{36} to restor the Hermiticity of these operators as was considered in
 the  paper of Fring and his colleagues \cite{1}. This situation is currently under investigation and is the goal of  my next work. Finally,
referring to Fring \textit{et al}'s work and this one, the  position dependent noncommutative space-time can be generalized as 
\begin{eqnarray}
[\hat X,\hat Y]=i\theta f(\hat Y),\quad [\hat X,\hat P_x]=i\hbar f(\hat Y), \quad 
[\hat Y,\hat P_y]=i\hbar f(\hat Y),
\end{eqnarray}
where $f$ is called function of deformation and we assume that it is strictly
positive ($f > 0$). Based on these equations, one can ask the question:
For what function of deformation $f$ there exists  nonzero minimal uncertainties or maximal uncertainties?

\section*{Acknowledgments}
I would like to thank the referees for giving such constructive comments
which considerably improved the quality of the paper.

\section*{Appendix:\quad Jacoby Identities}
In this appendix, we prove all the possible Jacoby identities of the proposition $3.1$
\begin{eqnarray}
 [[\hat X,\hat Y], \hat X]+  [[\hat Y,\hat X],\hat X]+ [[\hat X,\hat X],\hat Y]&=&0,\\
 {[[\hat X,\hat Y]}, \hat Y]+  [[\hat Y,\hat Y],\hat X]+ [[\hat Y,\hat X],\hat Y]&=&0,\\
 {[[\hat X,\hat Y]}, \hat P_x]+  [[\hat Y,\hat P_x],\hat X]+ [[\hat P_x,\hat X],\hat Y]&=&0\\
 {[[\hat X,\hat Y]}, \hat P_y]+  [[\hat Y,\hat P_y],\hat X]+ [[\hat P_y,\hat X],\hat Y]&=&0
 \end{eqnarray}
 \begin{eqnarray}
  {[[\hat X,\hat P_x]}, \hat Y]+  [[\hat P_x,\hat Y],\hat X]+ [[\hat Y,\hat X],\hat P_x]&=&0\\
 {[[\hat X,\hat P_x]}, \hat X]+  [[\hat P_x,\hat X],\hat X]+ [[\hat X,\hat X],\hat P_x]&=&0\\
 {[[\hat X,\hat P_x]}, \hat P_x]+  [[\hat P_x,\hat P_x],\hat X]+ [[\hat P_x,\hat X],\hat P_x]&=&0\\
  {[[\hat X,\hat P_x]}, \hat P_y]+  [[\hat P_x,\hat P_y],\hat X]+ [[\hat P_y,\hat X],\hat P_x]&=&0
\end{eqnarray}
 \begin{eqnarray}
   {[[\hat Y,\hat P_y]}, \hat X]+  [[\hat P_y,\hat X],\hat Y]+ [[\hat X,\hat Y],\hat P_y]&=&0\\
    {[[\hat Y,\hat P_y]}, \hat Y]+  [[\hat P_y,\hat Y],\hat Y]+ [[\hat Y,\hat Y],\hat P_y]&=&0\\
      {[[\hat Y,\hat P_y]}, \hat P_x]+  [[\hat P_y,\hat P_x],\hat Y]+ [[\hat P_x,\hat Y],\hat P_y]&=&0\\
   {[[\hat Y,\hat P_y]}, \hat P_y]+  [[\hat P_y,\hat P_y],\hat Y]+ [[\hat P_y,\hat Y],\hat P_y]&=&0\\
 \end{eqnarray}
 \begin{eqnarray}
   {[[\hat P_x,\hat P_y]}, \hat X]+  [[\hat P_y,\hat X],\hat P_x]+ [[\hat X,\hat P_x],\hat P_y]&=&0\\
     {[[\hat P_x,\hat P_y]}, \hat Y]+  [[\hat P_y,\hat Y],\hat P_x]+ [[\hat Y,\hat P_x],\hat P_y]&=&0\\
      {[[\hat P_x,\hat P_y]}, \hat P_x]+  [[\hat P_y,\hat P_x],\hat P_x]+ [[\hat P_x,\hat P_x],\hat P_y]&=&0\\
       {[[\hat P_x,\hat P_y]}, \hat P_y]+  [[\hat P_y,\hat P_y],\hat P_x]+ [[\hat P_y,\hat P_x],\hat P_y]&=&0\\
 \end{eqnarray}
 \begin{eqnarray}
   {[[\hat Y,\hat P_x]}, \hat X]+  [[\hat P_x,\hat X],\hat P_y]+ [[\hat X,\hat P_y],\hat P_x]&=&0\\
    {[[\hat Y,\hat P_x]}, \hat Y]+  [[\hat P_x,\hat Y],\hat P_y]+ [[\hat Y,\hat P_y],\hat P_x]&=&0,\\
     {[[\hat Y,\hat P_x]}, \hat P_x]+  [[\hat P_x,\hat P_x],\hat Y]+ [[\hat P_x,\hat Y],\hat P_x]&=&0,\\
      {[[\hat Y,\hat P_x]}, \hat P_y]+  [[\hat P_x,\hat P_y],\hat Y]+ [[\hat P_y,\hat Y],\hat P_x]&=&0
 \end{eqnarray}
\begin{eqnarray}
  {[[\hat X,\hat P_y]}, \hat X]+  [[\hat P_y,\hat X],\hat X]+ [[\hat X,\hat X],\hat P_y]&=&0\\
  {[[\hat X,\hat P_y]}, \hat Y]+  [[\hat P_y,\hat Y],\hat X]+ [[\hat Y,\hat X],\hat P_y]&=&0\\
  {[[\hat X,\hat P_y]}, \hat P_x]+  [[\hat P_y,\hat P_x],\hat X]+ [[\hat P_x,\hat X],\hat P_y]&=&0\\
 {[[\hat X,\hat P_y]}, \hat P_y]+  [[\hat P_y,\hat P_y],\hat X]+ [[\hat P_y,\hat X],\hat P_y]&=&0
\end{eqnarray}


\begin{thebibliography}{99}
\bibitem{1}  A. Fring,  L. Gouba   and F. Scholtz, {\it Strings from position-dependent noncommutativity},  J. Phys. A: Math. Theor. {\bf 43} (2010)  345401
\bibitem{2}   D. Amati, M. Ciafaloni and  G. Veneziano, 
               {\it Can Space-Time Be Probed Below the String Size?},
               Phys.Lett. B {\bfseries 216} (1989) 41-47
               
\bibitem{3}    D. Amati, M. Ciafaloni and  G. Veneziano,
              {\it Higher Order Gravitational Deflection and Soft Bremsstrahlung    in  Planckian Energy Superstring Collisions}, Nucl. Phys. B {\bfseries 347}, (1990) 550-580
             
\bibitem{4}    A. Kempf, G. Mangano and R. Mann, {\it Hilbert space representation
of the minitial length uncertainty relation},
 Phys. Rev. D {\bfseries 52}  (1995) 1108
\bibitem{5}  A. Kempf, {\it Uncertainty relation group symmetry in quantum mechanics with quantum }, J. Math. Phys. {\bf 35} (1994) 4483 
\bibitem{6} A. Kempf, {\it Non-pointlike particles in harmonic oscillators}, 
              J. Phys. A: Math. Gen. {\bf 30} (1997) 2093–2101

\bibitem{7}  F. Scardigli, {\it Generalized uncertainty principle in quantum gravity from micro-black hole gedanken experiment}, Phys. Lett. B, {\bf 452} (1999) 39-44
\bibitem{8}  F. Scardigli and  R. Casadio, {\it Generalized uncertainty principle, extra dimensions and holography},  Class.Quant.Grav. {\bf 20} (2003) 3915-3926

\bibitem{9} G. Lambiase, F. Scardigli, {\it Lorentz violation and generalized uncertainty principle}, Phys.Rev. D {\bf 97} (2018) 075003 

\bibitem{10}  V. Todorinov,  P. Bosso and S. Das {\it Relativistic Generalized Uncertainty Principle} arXiv 1810.11761v1 [gr-qc] 28 oct 2018

\bibitem{11} A. Kempf {\it Quantum Field Theory with Nonzero Minimal Uncertainties in Positions and Momenta},  Czech J Phys {\bf 44} (1994) 1041   
\bibitem{12} F. Scardigli, G. Lambiase, E. Vagenas {\it GUP parameter from quantum corrections to the Newtonian potential},  Phys. Lett. B {\bf 767} (2017) 242 
\bibitem{13} T. Kanazawa, G. Lambiase,  G. Vilasi and  A. Yoshioka, {\it Noncommutative Schwarzschild geometry and generalized uncertainty principle}, Eur. Phys. J. C
 {\bf 79} (2019) 95
\bibitem{14} Yen Chin Ong, {\it Generalized uncertainty principle,
black holes, and white dwarfs:a tale of two infinities}, Journal of Cosmology and
Astroparticle Physics doi:10.1088/1475-7516/2018/09/015

\bibitem{15}    A. Ali, S. Das and E. Vegenas, {\it Discreteness of space from the generalized uncertainty principle},
 Phys. Lett. B {\bfseries 678} (2009) 497-499
 
\bibitem{16}    P. Pedram, {\it A higher order GUP with minimal length uncertainty and maximal momentum},
 Phys. Lett. B {\bfseries 714} (2012) 317-323 
 
\bibitem{17}    P. Pedram, {\it A higher order GUP with minimal length uncertainty and maximal momentum II: Applications},
 Phys. Lett. B {\bfseries 718} (2012) 638-645 
 
\bibitem{18}   Y. Sabir and K. Nouicer, {\it Phase transitions of a GUP-corrected Schwarzschild black hole within isothermal cavities},
 Class. Quant. Grav {\bfseries 29} (2012) 215015 



\bibitem{19}  S. Dey, A. Fring  and L. Gouba,  {\it $\mathcal{PT}$-symmetric noncommutative spaces with minimal volume uncertainty relations}, J. Phys. A: Math. Theor. {\bf 45} (2014) 385302
\bibitem{20} A.Fring, L. Gouba  and  B. Bagchi, {\it Minimal areas from q-deformed oscillator algebras}, J. Phys. A: Math. Theor. {\bf 43}  (2010) 425202
\bibitem{21}  S. Alavi  and S. Abbaspour,   {\it Dynamical noncommutative quantum mechanics} J. Phys. A: Math. Theor. {\bf 47} (2014) 045303

\bibitem{22} S. Dey and A. Fring,  {\it The two dimensional harmonic oscillator on a
noncommutative space with minimal uncertainties},  	Acta Polytechnica 53 (2013) 268-276
\bibitem{23}   L. Lawson, L. Gouba and G. Avossevou, {\it Two-dimensional noncommutative gravitational quantum well}, J. Phys A: Math. Theor {\bfseries 50} (2017)   475202 

\bibitem{24} G. Camelia, {\it Relativity in space-times with short-distance structure governed by an observer-independent (Planckian) length scale}       Int. J. Mod. Phys. D {\bf 11}, (2000) 35 
\bibitem{25} G. Camelia, {\it Doubly Special Relativity}, Nature {\bf 418}, (2002) 34 
\bibitem{26} K. Nozari and A. Etemadi, {Minimal length, maximal momentum and Hilbert space representation of quantum mechanics} Phys. Rev. D {\bf 85}, (2012) 104029



\bibitem{27} S. Kresic-Juric, S. Meljanac, M. Stojic,
{\it Covariant realizations of kappa-deformed space},
Eur.Phys.J.C  {\bf 51} (2007) 229-240

 \bibitem{28} E. Harikumar, T. Juric, S. Meljanac,
 {\it Electrodynamics on $\kappa$-Minkowski space-time},
Phys. Rev. D {\bf 84}  (2011) 085020 

\bibitem{29} T. Juric, S. Meljanac, R. Strajn 
{\it  Twists, realizations and Hopf algebroid structure of kappa-deformed phase space },
Int. J. of Modern Physics A {\bf 29}  (2014) 1450022

\bibitem {30} T. Juric, S. Meljanac, D. Pikutic,
{\it Realizations of $\kappa$-Minkowski space, Drinfeld twists and related symmetry algebras}, Eur. Phys. J. C {\bf 75} (2015) 528

\bibitem{31} T. Juric, S. Meljanac, D. Pikutic, R. Strajn
{\it Toward the classification of differential calculi on  $\kappa$-Minkowski space and related field theories},
JHEP {\bf 1507} (2015) 055




\bibitem{32} F. Scholtz, L. Gouba, A. Hafver and C. Rohwer, {\it Formulation, interpretation and application of non-commutative quantum mechanics}, J. Phys. A: Math. Theor. {\bf 42}, (2009) 175303

        
\bibitem{33} O. Bertolami, J. Rosa , C. de Arag, P. Castorina   and D. Zappalà,               
              {\it Noncommutative gravitational quantum well}, Phys. Rev. D {\bf 72}, (2005) 025010 
              

\bibitem{34}  P. Pedram, {\it On the modification of Hamiltonians’ spectrum in gravitational quantum mechanics},  EPL. {\bf 89},  (2010) 50008

\bibitem{35}  P. Pedram {\it New Approach to Nonperturbative Quantum Mechanics with Minimal Length Uncertainty}, Phys. Rev. D, {\bf 85} (2012) 02401

\bibitem{36}  F. J. Dyson, {\it Thermodynamic Behavior of an Ideal Ferromagnet}, Phys. Rev. 102 (1956)  1230–1244.

             

  
              
     

\end{thebibliography}
\end{document}